\documentclass[aps,prl,twocolumn,showpacs]{revtex4-1}
\usepackage{graphicx,amsmath,color}

\begin{document}

\title{Direct experimental certification of quantum non-Gaussian character \\ and Wigner function negativity of single-photon detectors}

\author{Josef Hlou\v{s}ek}
\email{hlousek@optics.upol.cz}
\affiliation{Department of Optics, Palack\' y University, 17. listopadu 1192/12, 77146 Olomouc, Czech Republic}

\author{Miroslav Je\v{z}ek}
\email{jezek@optics.upol.cz}
\affiliation{Department of Optics, Palack\' y University, 17. listopadu 1192/12, 77146 Olomouc, Czech  Republic}

\author{Jarom\'{\i}r Fiur\'{a}\v{s}ek}
\email{fiurasek@optics.upol.cz}
\affiliation{Department of Optics, Palack\' y University, 17. listopadu 1192/12, 77146 Olomouc, Czech  Republic}

\begin{abstract}
Highly nonclassical character of optical quantum detectors, such as single-photon detectors, is essential for preparation of quantum states of light and a vast majority of applications in quantum metrology and quantum information processing. Therefore, it is both fundamentally interesting and practically relevant to investigate the nonclassical features of optical quantum measurements. Here we propose and experimentally demonstrate a procedure for direct certification of quantum non-Gaussianity and Wigner function negativity, two crucial nonclassicality levels, of photonic quantum detectors. Remarkably, we characterize the highly nonclassical properties of the detector by probing it with only two classical thermal states and a vacuum state. We experimentally demonstrate the quantum non-Gaussianity of single-photon avalanche diode even under the presence of background noise, and we also certify the negativity of the Wigner function of this detector. Our results open the way for direct benchmarking of photonic quantum detectors with a few measurements on classical states.
\end{abstract}

\maketitle

Nonclassical states of light are of fundamental importance in quantum optics, optical quantum communication, quantum information processing, and quantum metrology. The nonclassical states of optical fields are commonly defined as those whose Glauber-Sudarshan representation does not satisfy properties of an ordinary probability distribution. An important subclass of nonclassical states is represented by states with negative Wigner function \cite{Lvovsky2001,Zavatta2004,Ourjoumtsev2006,Nielsen2006,Wakui2007,Bertet2002}. Recently, another interesting sub-class of nonclassical states has been proposed, termed quantum non-Gaussian states \cite{Filip2011}. These states are defined as states whose density matrix cannot be expressed as a convex mixture of Gaussian states. Preparation of quantum non-Gaussian states thus requires nonlinear interaction or detection beyond the class of Gaussian operations that comprise interference in passive linear optical interferometers, quadrature squeezing, and homodyne detection.
While every state with negative Wigner function is a quantum non-Gaussian state, the class of quantum non-Gaussian states is strictly larger and contains also states with positive Wigner function. During recent years, the quantum non-Gaussian states have been the subject of intensive research \cite{Filip2011,Lachman2013,Paris2013,Hughes2014,Park2015,Lachman2016,Park2017,Schleich2018,Kuhn2018,Jezek2011,Jezek2012squeezed,Straka2014,Predojevic2014,Straka2018,Lachman2018,Treps2019,Song2013,Hughes2014,ParisFerraro2018,Takagi2018,Park2019,Park2019B}. Several criteria and witnesses for detection of quantum non-Gaussian states have been established \cite{Filip2011,Lachman2013,Paris2013,Hughes2014,Park2015,Lachman2016,Park2017,Schleich2018,Kuhn2018}, and the quantum non-Gaussian character of various sources of nonclassical light has been demonstrated experimentally \cite{Jezek2011,Jezek2012squeezed,Straka2014,Predojevic2014,Straka2018,Lachman2018,Treps2019}.

\begin{figure}[b!]
	\centering
	\includegraphics[width=1.0\linewidth]{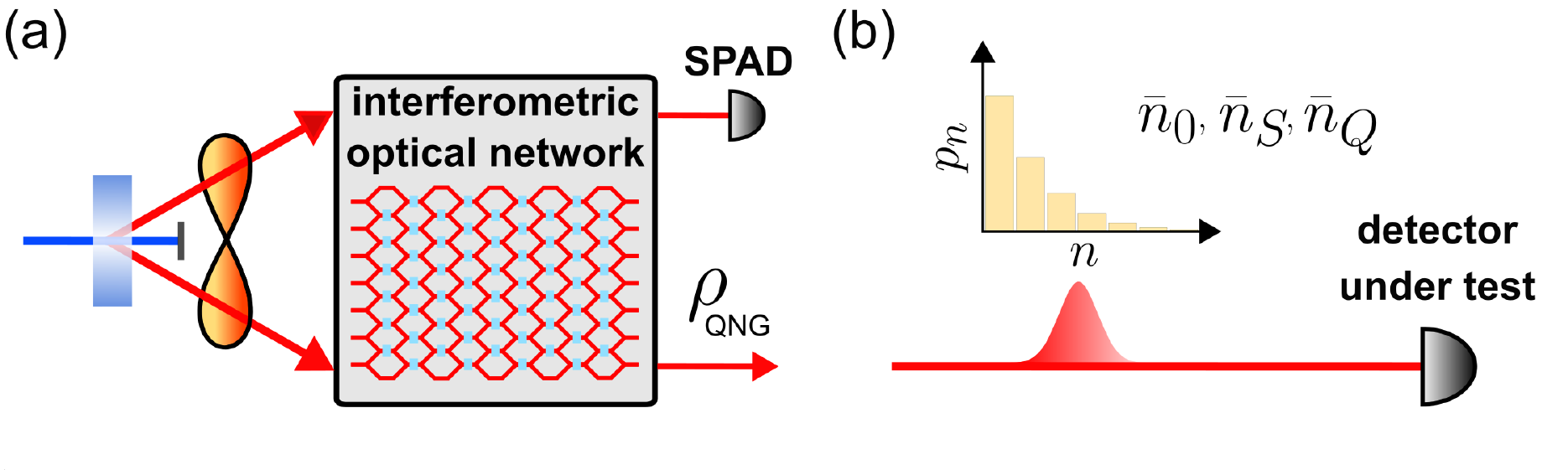}
	\caption{(a) Indirect certification of quantum non-Gaussian character of a single-photon detector, namely a single-photon avalanche diode (SPAD). A suitable multimode nonclassical Gaussian state is prepared using a parametric nonlinear process and an inteferometric network. A measurement with the single photon detector is performed on some of the output modes. Quantum non-Gaussianity of the heralded state certifies quantum non-Gaussianity of the detector. (b) Direct certification of quantum non-Gaussian character or negativity of the Wigner of a detector by probing it with two classical thermal states and a vacuum state.}
	\label{fig_1}
\end{figure}

The most common way to generate a quantum non-Gaussian state of light is to first generate a suitable multimode nonclassical Gaussian state, perform measurements with single-photon detectors on some of its modes and condition on photon detection \cite{Lvovsky2001,Zavatta2004,Ourjoumtsev2006,Nielsen2006,Wakui2007,Jezek2011,Jezek2012squeezed,Treps2019,Lvovsky2020arxiv}. This is schematically illustrated in Fig.~\ref{fig_1} (a). The subsequent certification of quantum non-Gaussian character of the conditionally generated state can then be interpreted as an indirect certification of quantum non-Gaussian character of the measurement performed on the auxiliary modes. For instance, when conditioning on click of a single-photon avalanche diode, the POVM element whose quantum non-Gaussian character is effectively tested and certified reads $\Pi=I-|0\rangle\langle0 |$ for an ideal detector with unit efficiency, or more realistically
\begin{equation}
\Pi=\sum _{n=0}^\infty\left [1-(1-\eta)^n\right]|n\rangle\langle n|
\end{equation}
for a detector with efficiency $\eta$.

In the present work we address the question whether the quantum non-Gaussian character of a quantum detector can be characterized more directly by performing measurements on suitably chosen probe states. We answer this question in affirmative and show that, interestingly, it may suffice to probe the detector with the vacuum state and two classical thermal states of different temperatures, see Fig.~\ref{fig_1} (b). Our procedure thus requires much less probe states than full quantum detector tomography \cite{Luis1999,Fiurasek2001,Lundeen2009,DAuria2011,Brida2012,Natarajan2013,Smith2014,Ma2016,Grandi2017,Izumi2020}. In our approach, the role of measurement and state is reversed with respect to the ordinary certification of quantum non-Gaussian character of quantum states. One specific feature of certification of quantum non-Gaussian character of some POVM element $\Pi$ is that $\mathrm{Tr}[\Pi]$ need not be finite. We show that this obstacle can be circumvented by characterizing a suitably regularized POVM.

We begin with a short recapitulation of the quantum non-Gaussianity criterion that will be utilized in our work and adapted for direct characterization of quantum non-Gaussianity of quantum measurements. Consider a quantum state $\rho$ and let $p_0=\langle 0|\rho|0\rangle$ denote the probability of vacuum state in $\rho$. Let $\mathcal{L}$ denote a lossy channel with $50\%$ transmittance, and define $q_0=\langle 0|\mathcal{L}(\rho)|0\rangle$ as the probability of vacuum state  at the output of the lossy channel. It holds that
\begin{equation}
q_0=\sum_{n=0}^\infty \frac{p_n}{2^n} ,
\label{q0definition}
\end{equation}
where $p_n=\langle n|\rho |n\rangle$ is the photon number distribution of state $\rho$. As shown in Ref. \cite{Lachman2013}, a quantum non-Gaussianity criterion can be formulated in terms of $q_0$ and $p_0$. Specifically, the state is proved to be quantum non-Gaussian if $q_0$ exceeds certain threshold that depends on $p_0$. The dependence of maximum $q_0$ achievable for a given $p_0$ with Gaussian states and their mixtures can be expressed analytically in a parametric form \cite{Lachman2013},
\begin{eqnarray}
p_0(V)&=&\frac{2\sqrt{V}}{V+1}\exp\left[ -\frac{(1-V)(3+V)}{2V(3V+1)}  \right], \nonumber \\
q_0(V)&=&\frac{4\sqrt{V}}{\sqrt{(V+3)(3V+1)}}\exp\left[- \frac{1-V^2}{2V(3V+1)}\right],
\label{p0q0criterion}
\end{eqnarray}
where $V\in(0,1]$.

We now turn our attention to certification of quantum non-Gaussianity of a quantum measurement, and we specifically consider a particular measurement outcome associated with a POVM element $\Pi$. We can treat $\Pi$ as a non-normalized density matrix.  Assuming that $\mathrm{Tr}[\Pi]$ is finite, we could introduce normalized operator $\rho_\Pi=\Pi/\mathrm{Tr}[\Pi]$, treat it as an equivalent of a density matrix and define
\begin{equation}
P_0=\frac{1}{\mathrm{Tr}[\Pi]}\langle 0|\Pi|0\rangle, \qquad
Q_0=\frac{1}{\mathrm{Tr}[\Pi]}\sum_{n=0}^\infty \frac{1}{2^n}\langle n|\Pi|n\rangle.
\label{P0Q0}
\end{equation}
If we knew $Q_0$ and $P_0$ then we could apply the above described quantum non-Gaussianity criterion based on Eq. (\ref{p0q0criterion})
to verify the quantum non-Gaussian character of the POVM element $\Pi$. However, $\mathrm{Tr}[\Pi]$ can be infinite, and even if finite, it could be difficult to measure.

We circumvent this obstacle by considering a regularized version of the POVM element
\begin{equation}
\tilde{\Pi}=W\Pi W^\dagger,
\label{Pitilde}
\end{equation}
where
\begin{equation}
W=\sum_{n=0}^\infty \nu^n |n\rangle\langle n|.
\end{equation}
The transformation (\ref{Pitilde}) represents a noiseless quantum attenuation \cite{Micuda2012} with factor $0<\nu<1$. The noiseless quantum attenuation is a conditional Gaussian quantum operation that can be applied to a quantum state of an optical mode by sending the state through a beam splitter with amplitude transmittance $\nu$, whose auxiliary input port is prepared in vacuum state and whose auxiliary output port is projected onto vacuum state. If $\rho$ is a Gaussian state or a mixture of Gaussian states, then also $W\rho W^\dagger$ is a mixture of Gaussian states. Therefore, if we find that $W\rho W^\dagger$ is quantum non-Gaussian, then also $\rho$ must have been quantum non-Gaussian state. This observation can be straightforwardly extended to quantum measurements. In particular, if we find that $\tilde{\Pi}$ is quantum non-Gaussian, then also the original POVM element $\Pi$ must have been quantum non-Gaussian. We note that the quantum state $\rho=\tilde{\Pi}^{\mathrm{T}}/\mathrm{Tr}[\tilde{\Pi}]$, where the transposition is performed in the Fock basis, can be conditionally generated by preparing a Gaussian two-mode squeezed vacuum state 
\begin{equation}
|\Psi\rangle_{AB}=\sqrt{1-\nu^2}\sum_{n=0}^\infty \nu^n |n,n\rangle_{AB}
\end{equation}
and performing the quantum measurement on mode B. Measurement outcome associated with the POVM element $\Pi$ then heralds preparation of mode A in state $\rho_{\tilde{\Pi}}$. This shows that the regularization (\ref{Pitilde}) of the POVM element $\Pi$ has a direct experimental relevance and characterization of $\tilde{\Pi}$  specifies the ultimately achievable properties of the conditionally prepared state $\rho_{\tilde{\Pi}}$,
when all other aspects of the state preparation are perfect.

Considering $\tilde{\Pi}$ instead of $\Pi$, we find that its trace is always well defined and finite,
\begin{equation}
S=\mathrm{Tr}[\tilde{\Pi}]
=\sum_{n=0}^\infty \nu^{2n}\langle n|\Pi|n\rangle.
\end{equation}
In fact, the quantity $S$ can be exprimentally determined by probing the detector with a thermal state. Recall that the density matrix of a thermal state with mean photon number $\bar{n}$ reads
\begin{equation}
\rho_{\mathrm{th}}(\bar{n})=\frac{1}{\bar{n}+1}\sum_{n=0}^\infty \left( \frac{\bar{n}}{\bar{n}+1}\right)^n |n\rangle \langle n|.
\end{equation}
By choosing $\bar{n}_S=\nu^2/(1-\nu^2)$ we get $\nu^2=\bar{n}_S/(\bar{n}_S+1)$ and
\begin{equation}
S=\frac{1}{1-\nu^2}\mathrm{Tr}[\rho_{\mathrm{th}}(\bar{n}_S) \Pi],
\end{equation}
where $\mathrm{Tr}[\rho_{\mathrm{th}}(\bar{n}_S)\Pi]$ represents the probability of outcome $\Pi$ when probing the detector with thermal state $\rho_{\mathrm{th}}(\bar{n}_S)$.

We can define the vacuum probabilities $\tilde{P}_0$ and $\tilde{Q}_0$ for the normalized operator $\rho_{\tilde{\Pi}}=\tilde{\Pi}/\mathrm{Tr}[\tilde{\Pi}]$ corresponding to the regularized POVM element $\tilde{\Pi}$
in full analogy to Eq. (\ref{P0Q0}),
\begin{eqnarray}
\tilde{P_0}&=&\frac{\langle 0|\tilde{\Pi} |0\rangle}{\mathrm{Tr}[\tilde{\Pi} ]}  =\frac{1}{S} \langle 0|\Pi|0\rangle, \nonumber \\[2mm]
\qquad
\tilde{Q}_0&=&\frac{1}{\mathrm{Tr}[\tilde{\Pi}]} \sum_{n=0}^\infty \frac{1}{2^n} \langle n|\tilde{\Pi}|n\rangle=\frac{1}{S}\sum_{n=0}^\infty \frac{\nu^{2n}}{2^n} \langle n| \Pi |n\rangle. \nonumber
\end{eqnarray}
It follows that $\Tilde{P}_0$ can be estimated by probing the detector with the vacuum state and $\tilde{Q}_0$ can be determined by probing the detector with thermal state with mean photon number $\bar{n}_Q=\nu^2/(2-\nu^2)$, which follows from $\frac{\nu^2}{2}=\bar{n}_Q/(\bar{n}_Q+1)$.
In terms of the mean numbers of thermal photons, the probabilities $\tilde{P}_0$ and $\tilde{Q}_0$ can be expressed as
\begin{equation}
\tilde{P}_0\!=\!\frac{1}{\bar{n}_S\!+\!1}\frac{\mathrm{Tr}[\Pi \,\rho_{\mathrm{th}}(0)]}{\mathrm{Tr}[\Pi \, \rho_{\mathrm{th}}(\bar{n}_S)]},\,
\tilde{Q}_0\!=\!\frac{\bar{n}_Q\!+\!1}{\bar{n}_S\!+\!1} \frac{\mathrm{Tr}[\Pi\,\rho_{\mathrm{th}}(\bar{n}_Q)]}{\mathrm{Tr}[\Pi\,\rho_{\mathrm{th}}(\bar{n}_S)]},\,
\label{PQnbar}
\end{equation}
where
\begin{equation}
\bar{n}_Q=\frac{\bar{n}_S}{\bar{n}_S+2}.
\end{equation}

Let us now consider a single photon avalanche diode (SPAD) with detection efficiency $\eta$ and probability of dark counts $R_D$.  Here $R_D$ is the probability that the detector clicks when a vacuum state is injected into the detected signal mode. We can describe the click outcome of such detector with the following POVM element,
\begin{equation}
\Pi_{\mathrm{SPAD}}=\sum _{n=0}^\infty\left [1-(1-R_D)(1-\eta)^n\right]|n\rangle\langle n|.
\end{equation}
 After some algebra, we obtain
 \begin{equation}
 \tilde{P}_0= \frac{R_D(1-\nu^2)(1-\nu^2+\nu^2\eta)}{R_D(1-\nu^2)+\nu^2\eta}
 \label{P0model}
 \end{equation}
and
\begin{equation}
\tilde{Q}_0=2\frac{R_D(2-\nu^2)+\nu^2\eta}{R_D(1-\nu^2)+\nu^2\eta} \times \frac{(1-\nu^2)(1-\nu^2+\nu^2\eta)}{(2-\nu^2)(2-\nu^2+\nu^2 \eta)}.
\label{Q0model}
\end{equation}

In the experiment, the mean photon numbers of thermal states will be estimated and calibrated with some uncertainty specified by a confidence interval,
$\bar{n}_j\in [\bar{n}_j^{-},\bar{n}_j^{+}]$. We now show how to take this uncertainty into account and obtain suitable bounds on $\tilde{P}_0$ and $\tilde{Q}_0$ for which the quantum non-Gaussianity criterion remains applicable. First we exploit the convexity of the set $\mathcal{G}$ of probability pairs $[p_0,q_0]$ that can be obtained from Gaussian states and their mixtures. Since $[0,0]$ is an extremal point of $\mathcal{G}$, it holds that if $[p_0,q_0]\in \mathcal{G}$, then also $[xp_0,xq_0]\in \mathcal{G}$, where $0\leq x \leq 1$. Conversely, if $[xp_0,xq_0 ] \notin\ \mathcal{G}$ then also $[p_0, q_0] \notin \mathcal{G}$. To account for the uncertainty in determination of $\bar{n}_S$ we can conservatively apply the quantum non-Gaussianity test to $[x \tilde{P}_0,x \tilde{Q}_0]$ instead of $[\tilde{P}_0,\tilde{Q}_0]$, where $x=\frac{\bar{n}_S+1}{\bar{n}_S^{+}+1}$. Practically, this means that we replace $\bar{n}_S$ with the upper bound $\bar{n}_S^{+}$ in the denominators of prefactors in Eq. (\ref{PQnbar}), and consider the probabilities
\begin{equation}
\tilde{P}_0^\prime\!=\!\frac{1}{\bar{n}_S^{+}\!+\!1} \frac{\mathrm{Tr}[\Pi \,\rho_{\mathrm{th}}(0)]}{\mathrm{Tr}[\Pi \, \rho_{\mathrm{th}}(\bar{n}_S)]},\tilde{Q}_0^\prime\!=\! \frac{\bar{n}_Q\!+\!1}{\bar{n}_S^{+}\!+\!1} \frac{\mathrm{Tr}[\Pi \,\rho_{\mathrm{th}}(\bar{n}_Q)]}{\mathrm{Tr}[\Pi \, \rho_{\mathrm{th}}(\bar{n}_S)]}.
\label{PQprimed}
\end{equation}
The utilized quantum non-Gaussianity criterion certifies quantum non-Gaussianity when $\tilde{Q}_0'$ is larger than certain threshold that depends on $\tilde{P}_0'$. Since $\tilde{Q}_0'$ is an increasing function of $\bar{n}_Q$ we can avoid false positive certification of quantum non-Gaussianity due to uncertainty in $\bar{n}_Q$ calibration by utilizing a sufficiently low $\bar{n}_Q$ such that
\begin{equation}
\bar{n}_Q^{+}\leq \frac{\bar{n}_S^{-}}{\bar{n}_S^{-}+2}
\label{nQcondition}
\end{equation}
is satisfied. The condition (\ref{nQcondition}) ensures that for any pair of true values of the mean photon numbers $\bar{n}_S$ and $\bar{n}_Q$ from the confidence intervals the estimated $\tilde{Q}_0'$ will not exceed the value of $\tilde{Q}_0'$ that would be obtained for $\bar{n}_Q=\bar{n}_S/(\bar{n}_S+2)$.  To summarize, if one satisfies the condition (\ref{nQcondition}) and utilizes the above defined $\tilde{P}_0^\prime$ and $\tilde{Q}_0^{\prime}$ then one can certify quantum non-Gaussianity of the POVM element $\Pi$ even with uncertainty in calibration of the mean photon numbers $\bar{n}_S$ and $\bar{n}_Q$.

In addition to certification of quantum non-Gaussianity,  the probabilities $\tilde{P}_0$ and $\tilde{Q}_0$ can also be used to certify negativity of Wigner function that represents the characterized POVM element $\Pi$. Returning for a while back to $q_0$ and $p_0$, it follows from Eq. (\ref{q0definition}) that
\begin{equation}
q_0-p_0= \frac{p_1}{2}+\sum_{n=2}^\infty\frac{p_n}{2^n} \leq \frac{p_1}{2}+\frac{1}{4}\sum_{n=2}^\infty p_n.
\label{qpinequality}
\end{equation}
Since $\sum_{n=2}^\infty p_n=1-p_0-p_1$, we obtain from the inequality (\ref{qpinequality}) a lower bound on $p_1$,
\begin{equation}
p_1 \geq 4q_0-3p_0-1.
\label{p1bound}
\end{equation}
If $p_1>\frac{1}{2}$ then the Wigner function of state $\rho$ with photon number distribution $p_n$ is negative at the origin. Using inequality (\ref{p1bound}), the negativity of  Wigner function of an optical quantum state can be certified from a simple measurement with a balanced beam splitter and a pair of detectors, that is commonly used for measurement of the anticorrelation factor \cite{GrangierAspect1986,Jezek2011}. If we apply the condition (\ref{p1bound}) to the normalized operator $\rho_{\tilde{\Pi}}=\tilde{\Pi}/\mathrm{Tr}[\tilde{\Pi}]$, we find that the Wigner function of $\tilde{\Pi}$ must be negative at the origin of phase space if
\begin{equation}
4\tilde{Q}_0-3\tilde{P}_0-1 >\frac{1}{2}.
\label{QPWigner}
\end{equation}
Since the noiseless attenuation (\ref{Pitilde}) as a Gaussian operation preserves positivity of Wigner function of the POVM element, the inequality (\ref{QPWigner}) implies also the negativity of Wigner function of the original POVM element $\Pi$.

\begin{figure}[ht!]
	\centering
	\includegraphics[width=1.0\linewidth]{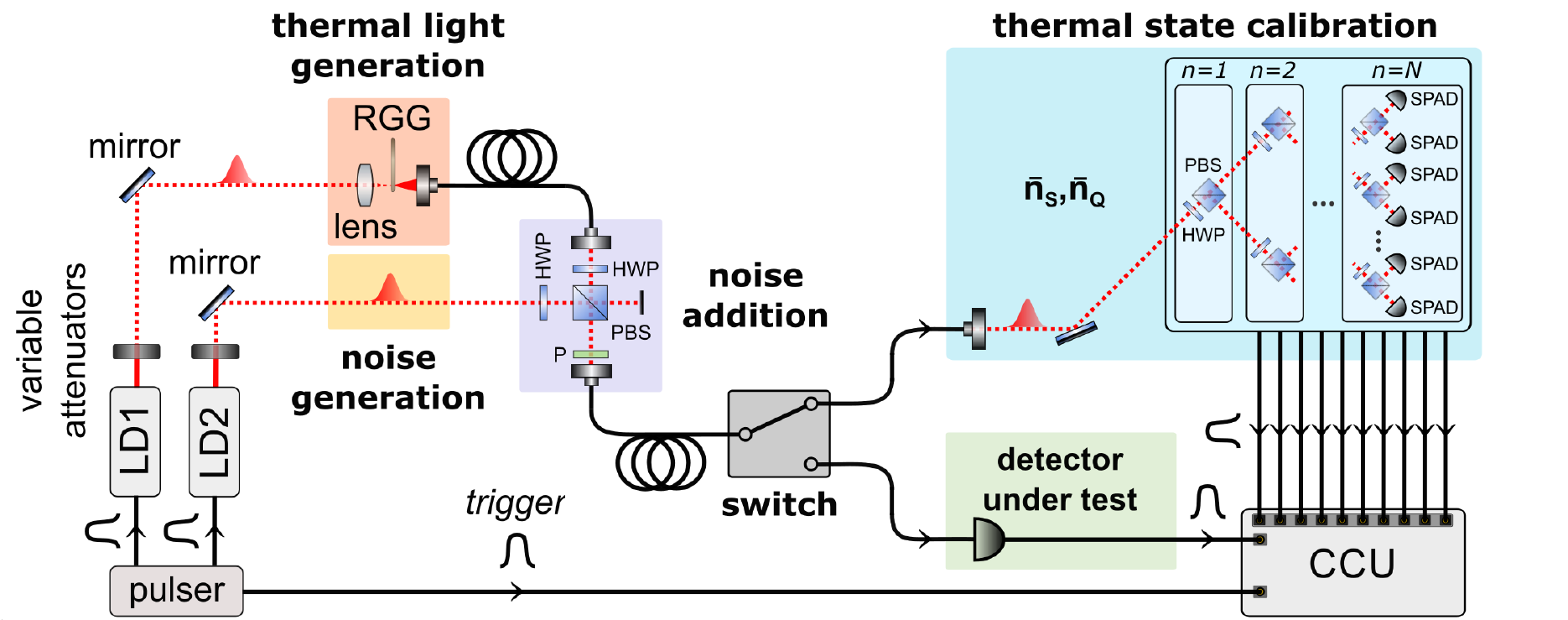}
	\caption{Experimental setup: the scheme includes preparation of pseudo-thermal state using a laser diode (LD1) and rotating ground glass (RGG); noise generation and addition using a laser diode (LD2), polarizing beam splitter (PBS), and polarizer (P); calibration of mean photon numbers of thermal states; detector under test; coincidence counting unit (CCU).}
	\label{fig_setup}
\end{figure}

The experimental setup for direct certification of quantum non-Gaussianity of a single-photon detector is shown in Fig.~\ref{fig_setup}. Nanosecond optical pulses with the central wavelength of 0.8 $\mu$m are produced by a gain-switched semiconductor laser diode driven by an electronic pulser at a repetition rate of 1 MHz. The pulser drives also an auxiliary laser diode emulating noise of the detector under test, and provides an electronic trigger signal. The train of coherent laser pulses is attenuated and focused on rotating ground glass (RGG); the output is collected by a single-mode fiber to produce pseudo-thermal light with Bose-Einstein distribution \cite{Spiller1964}. Let us stress that other single-mode thermal sources can be used here, such as direct intensity modulation \cite{Straka2018generator} or thermal emission from atomic ensembles \cite{Slodicka2018}. The photodistribution of the generated thermal light and its mean photon number $\bar{n}$ per pulse is verified by a spatially multiplexed photon-number-resolving detector (PNRD) \cite{Hlousek2019}. 

The PNRD consists of cascaded tunable beam splitters composed of a half-wave plate (HWP) and a polarizing beam splitter (PBS), which allow for accurate balancing of the output ports. The whole network works as a 1-to-$10$ splitter balanced with the absolute error below $0.3\%$. To measure the multiplexed signal we use SPADs with the typical efficiency ranging from 55 to 70\%, 250~ps timing jitter, and 25~ns recovery time. The total detection efficiency of 50(1)\% is determined based on measured transmittance of the network and SPAD efficiencies specified by the manufacturer. For independent verification of the efficiency an absolute method using correlated photons can be used \cite{Migdall2007,Eisenberg2018}. The electronic outputs of the SPADs are processed by a custom coincidence unit (CCU) while keeping the individual channels synchronized to the trigger. The emitter-coupled logic circuitry of the coincidence unit consists of fast comparators, delay lines and basic gates, and achieves 10 ps timing resolution and 20 ps overall jitter \cite{Hlousek2020ccu}. The single-run output of the unit is stored in flip-flop gates and is read by a microcontroller and sorted in a coincidence histogram for repeated measurements. For the presented experiment the histogram is reduced to just 11 elements corresponding to coincidence clicks of $D$ detectors, with $0 \leq D\leq 10$. The photon-number statistics of the optical signal is retrieved from the measured coincidence histogram using the expectation-maximization-entropy algorithm \cite{Hlousek2019}. The measured  photon-number statistics matches very well to the ideal Bose-Einstein statistics with the typical fidelity of 0.999, and the mean photon number $\bar{n}$ is accurately determined.

\begin{figure}[!h!]
	\centering
	\includegraphics[width=1.0\linewidth]{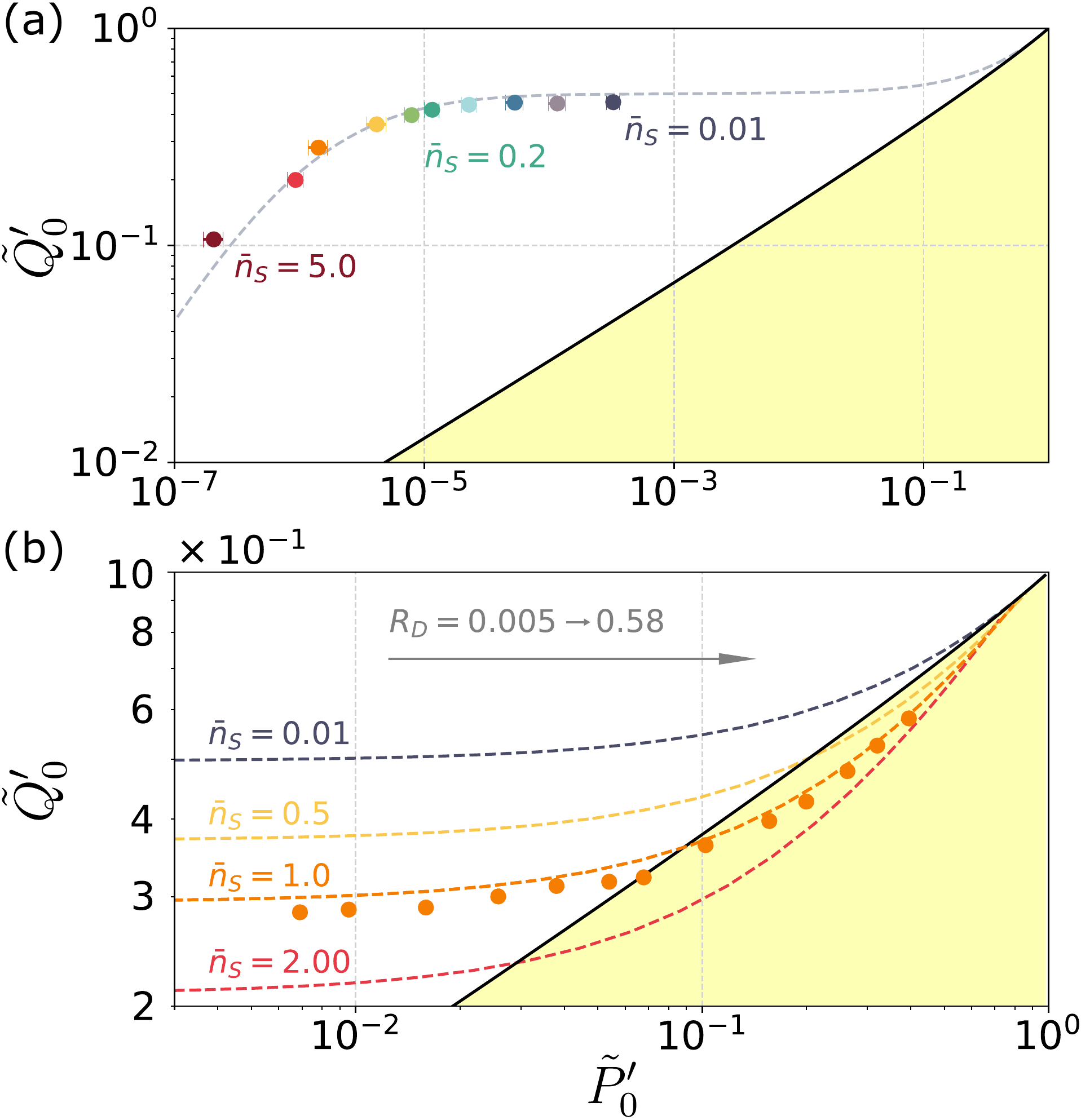}
	\caption{Quantum non-Gaussianity certification of single-photon avalanche diode for various mean photon numbers $\bar{n}_S$ of the probe thermal states (a). Quantum non-Gaussianity certification of single-photon avalanche diode for increased level of background noise and the particular value of the mean photon number $\bar{n}_S = 1.0$ of the probe thermal state (b). Measured data (markers) are compared to theoretical model (dashed curve). For reference, the theoretical curves in panel (b) are plotted for $4$ different values of $\bar{n}_S$. Error bars are smaller than data marker size. Quantum non-Gaussianity is certified for points lying outside the yellow area.}
	\label{fig_SPAD}
\end{figure}

\begin{figure}[!t!]
	\centering
	\includegraphics[width=1.0\linewidth]{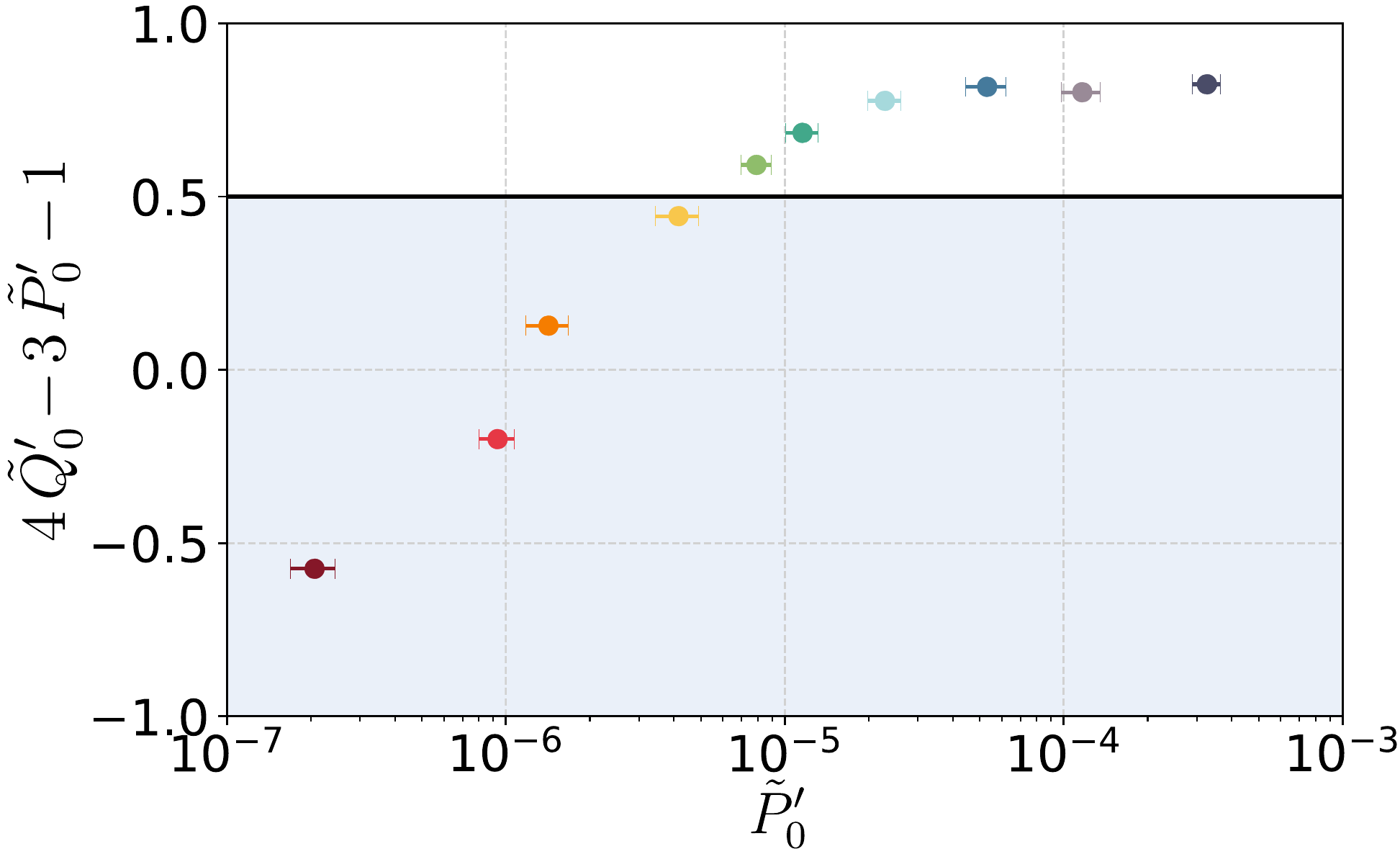}
	\caption{Certification of negativity of Wigner function of single-photon avalanche diode for various mean photon numbers $\bar{n}_S$ of the probe thermal states, color-coded similarly as in Fig.~\ref{fig_SPAD}(a).
		Vertical error bars are smaller than symbol size. Negativity of the Wigner function is certified when the threshold 0.5 is exceeded.}
	\label{fig_wigner}
\end{figure}

Characterized thermal light is then fed to the detector under test, which is, in our case, a SPAD module (SPCM-AQRH-14-FC manufactured by Excelitas) with the detection efficiency of 58\% specified by the manufacturer and measured dark count probability of $R_{\text{D}} = 1.44(8) \times 10^{-6}$.
The output electronic signal is again processed by the CCU and the number of detection events is evaluated together with the total number of the triger events, i.e. the number of input optical pulses. The relative detection frequency is acquired for vacuum state and thermal states with the mean photon number of $\bar{n}_Q$ and $\bar{n}_S$, which sample the ideal probabilities ${\mathrm{Tr}[\Pi\,\rho_{\mathrm{th}}(0)]}$, ${\mathrm{Tr}[\Pi\,\rho_{\mathrm{th}}(\bar{n}_Q)]}$, ${\mathrm{Tr}[\Pi\,\rho_{\mathrm{th}}(\bar{n}_S)]}$, respectively. The quantities $\tilde{P}_0^\prime$ and $\tilde{Q}_0^{\prime}$ are evaluated as in Eq.~(\ref{PQprimed}) and the resulting diagram is shown in Fig.~\ref{fig_SPAD}(a).
Ten pairs $[\tilde{P}_0^\prime,\,\tilde{Q}_0^{\prime}]$ are measured for the mean photon number $\bar{n}_S$ ranging from $0.0103 \pm 0.0006$ to $5.04 \pm 0.02$. The error bars represent one standard deviation. The measured data are compared to a theoretical plot based on the formulas (\ref{P0model},\ref{Q0model}). The results unequivocally certify quantum non-Gaussian character of the tested SPAD. The measured pairs $[\tilde{P}_0^\prime,\,\tilde{Q}_0^{\prime}]$ can be used also to certify negativity of Wigner function of the detector under test based on Eq.~(\ref{QPWigner}), see Fig.~\ref{fig_wigner}. We have unambiguously demonstrated Wigner function negativity of the characterized SPAD. The negativity threshold is surpassed by 132 standard deviations for $\bar{n}_S = 0.0103 \pm 0.0006$, with $4 \tilde{Q}_0^{\prime} - 3 \tilde{P}_0^\prime - 1 = 0.824(7)$.

Similarly to quantum non-Gaussianity depth defined for single-photon states \cite{Straka2014} we can explore resilience of quantum non-Gausianity of the single-photon detector to its imperfections, such as loss and noise. Dark counts of the detector and background light are of main interest in many applications, such as quantum key distribution \cite{Diamanti2016,Ling2017} or mapping and counting single-photon emitters \cite{Hell2015,Straka2018,Chekhova2018clusters}. To emulate background noise, we drive an auxiliary laser diode, and superimpose incoherently the resulting Poisson signal with the probe thermal light. Fig.~\ref{fig_SPAD}(b) shows the measured trajectory of the $[\tilde{P}_0^\prime,\,\tilde{Q}_0^{\prime}]$ pairs with increasing intensity of the noise for a fixed value of the probe mean photon number $\bar{n}_S = 1.0$. The data are compared to theoretical plots based on the formulas (\ref{P0model},\ref{Q0model}). We observe that when the noise exceeds certain threshold, the quantum non-Gaussian character of the detector cannot be certified any more with the applied criterion.

In conclusion, we have developed a method for direct verification of quantum non-Gaussianity of a quantum measurement that is based on probing the measurement apparatus with two classical thermal states and a vacuum state. The same experimental procedure also enables to certify negativity of Wigner function representing the characterized POVM element $\Pi$. We have experimentally demonstrated the direct certification of quantum non-Gaussianity of a quantum measurement for a single-photon avalanche photodiode and the observed experimental data are in excellent agreement with theoretical model. Our results extend the concept of quantum non-Gaussianity to quantum measurements and provide new simple and experimentally feasible tools to characterize highly non-classical features of quantum detectors with a few measurements. Our method can serve for fast and direct testing and benchmarking of various detectors including single-photon detectors, photon-number-resolving detectors, or even more complex detection schemes where the signal is preprocessed by some quantum operation such as squeezing or amplification before being detected. Since our method is wavelength-independent, it can be also utilized to characterize  detectors of microwave photons employed in cavity or circuit quantum electrodynamics \cite{Gleyzes2007,Schuster2007,Johnson2010,Royer2018,Kono2018,Opremcak2018,Besse2020}.

\begin{acknowledgments}
JF acknowledges support by the Czech Science Foundation under Grant No. 19-19722J. This work has received national funding from the MEYS and the funding from European Union's Horizon 2020 research and innovation framework programme under grant agreement No. 731473 (project 8C18002). Project Hyper-U-P-S has received funding from the QuantERA ERA-NET Cofund in Quantum Technologies implemented within the European Union's Horizon 2020 Programme.
\end{acknowledgments}


%

\end{document}